\documentclass[twocolumn,aps,prl,showpacs]{revtex4-1}
\usepackage[T1]{fontenc}
\usepackage[latin9]{inputenc}
\usepackage{textcomp}
\usepackage{amsmath}
\usepackage{amssymb}
\usepackage{stackrel}
\usepackage{graphicx}
\usepackage{esint}

\makeatletter

\@ifundefined{textcolor}{}
{%
 \definecolor{BLACK}{gray}{0}
 \definecolor{WHITE}{gray}{1}
 \definecolor{RED}{rgb}{1,0,0}
 \definecolor{GREEN}{rgb}{0,1,0}
 \definecolor{BLUE}{rgb}{0,0,1}
 \definecolor{CYAN}{cmyk}{1,0,0,0}
 \definecolor{MAGENTA}{cmyk}{0,1,0,0}
 \definecolor{YELLOW}{cmyk}{0,0,1,0}
}

\makeatother

\begin{document}

\title{Spin-exchange-induced exotic superfluids in a Bose-Fermi spinor mixture}

\author{Chuanzhou Zhu$^1$, Li Chen$^2$, Hui Hu$^3$, Xia-Ji Liu$^3$, and Han Pu$^1$}

\affiliation{ $^1${Department of Physics and Astronomy, and Rice Center for Quantum Materials,
	Rice University, Houston, Texas 77251-1892, USA}\\
$^2${Institute of Theoretical Physics and Department of Physics, State Key Laboratory of Quantum Optics and Quantum Optics Devices, and Collaborative Innovation Center of Extreme Optics, Shanxi University, Taiyuan 030006, China} \\
$^3${Centre for Quantum and Optical Science, Swinburne University of Technology, Melbourne 3122, Australia}}

\begin{abstract}
We consider a mixture of spin-1/2 bosons and fermions, where only the bosons are subjected to the spin-orbit coupling induced by Raman beams. The fermions, although not directly coupled to the Raman lasers, acquire an effective spin-orbit coupling through the spin-exchange interaction between the two species. Our calculation shows that this is a promising way of obtaining spin-orbit coupled Fermi gas without Raman-induced heating, where the long-sought topological Fermi superfluids and topological bands can be realized. Conversely, we find that the presence of fermions not only provides a new way to create the supersolid stripe phase of the bosons, but more strikingly it can also greatly increase the spatial period of the bosonic density stripes, and hence makes this phase directly observable in experiment. This system provides a new and practical platform to explore the physics of spin-orbit coupling, which possesses a dynamic nature through the interaction between the two species. 
\end{abstract}
\maketitle

{\em Introduction ---} In recent years, spin-orbit (SO) coupling in cold atoms
\cite{SOCBosonExp2,SOC2,SOC1,SOC5} has received tremendous attention. Experimental realization of SO coupled bosons \cite{SOCBosonExp3,SOCBosonExp4,SOCBosonExp5,SOCBosonExp55,SOCBosonExp6}
and fermions \cite{SOCFermiExp2,SOCFermiExp3,SOCFermiExp4,SOCFermiExp5,SOCFermiExp6,SOCFermiExp7} has been reported around the world. The interest in such systems is mainly due to the exotic phases induced by SO coupling in quantum gases. For example, SO coupled Bose-Einstein
condensates (BEC) can host a stripe
phase featuring spatially modulated density profiles \cite{SOCBosonTheory1,SOCBosonTheory2,SOCBosonTheory3,SOCBosonTheory4,SOCBosonTheory5,BFmixSOC1,BFmixSOC2}, which can be regarded as a supersolid \cite{SOCBosonExp6,liao}; whereas SO coupled attractive Fermi gas can become a topological superfluid, supporting Majorana edge states \cite{SOCFermionTopoBand1,SOCFermionTopoBand2,SOCFermionTopoBand3,SOCFermionTopoBand4,
SOCFermionTopoBand5,SOCFermionTopoBand6,SOCFermionTopoBand7,
SOCFermionMajorana1,SOCFermionMajorana2,SOCFermionMajorana3,SOCFermionMajorana4,SOCFermionMajorana5}.
However, the spatial period of the stripe phase is on the order of the optical wavelength, making its direct observation extremely challenging, although indirect evidences for the stripe phase have been reported in two seminal experiments~\cite{SOCBosonExp55,SOCBosonExp6}. Another serious experimental problem for realizing SO coupling in quantum gas concerns the heating due to the Raman beams. The Raman-induced heating is particularly severe for atomic species with small fine-structure splitting \cite{heating}. Wei and Mueller carefully analyzed the heating problems for all alkali-metal atoms \cite{heating}. According to their analysis, $^{40}$K and $^6$Li --- the two most commonly used fermionic species in cold atom experiments --- suffer greatly from such heating. This could explain why an SO coupled fermionic superfluid, despite its tremendous theoretical interest, has yet to be realized in experiment.

Here we consider a mixture of bosonic and fermionic superfluids, each of which is a spin-1/2 system. In addition to the density-density interactions, there exists an inter-species spin-exchange interaction. We assume that the condensate is subjected to the Raman-induced SO coupling, whereas the Fermi gas is not coupled by the Raman beams and hence is immune from the Raman-induced heating. The key observation of this Letter is that, through the spin-exchange interaction, the Fermi gas experiences a significant effective SO coupling. The interplay between the bosons and fermions leads to a variety of interesting quantum states, including a stripe phase in condensate with the spatial period much larger than the optical wavelength, and various topological phases for fermions. Note that a Bose-Bose spinor mixture has been realized in a recent experiment, and the associated effects of spin-exchange interaction were observed \cite{DJWang2} and theoretically analyzed \cite{BBmix}. In addition, several groups have created Bose-Fermi superfluid mixtures with scalar condensates  \cite{BFmixSuperfluid1,BFmixSuperfluid2,BFmixSuperfluid3,BFmixSuperfluid4}.

{\em Hamiltonian ---} The total
Hamiltonian of the system takes the form (we take $\hbar=1$)
\begin{equation}
\mathbf{\mathbf{H}}=\int dx\left(\Psi_{B}^{\dagger}h_{B}\Psi_{B}+\Psi_{F}^{\dagger}h_{F}\Psi_{F}\right)+\mathcal{\mathcal{G}}_{B}+\mathcal{\mathcal{G}}_{F}+\mathcal{\mathcal{G}}_{BF},\label{eq:H}
\end{equation}
where $\Psi_{B}\left(x\right)=\left[\psi_{B\uparrow}\left(x\right),\psi_{B\downarrow}\left(x\right)\right]^{\mathrm{T}}$
represents the mean-field wave function of the BEC, and $\Psi_{F}\left(x\right)=\left[\psi_{F\uparrow}\left(x\right),\psi_{F\downarrow}\left(x\right)\right]^{\mathrm{T}}$
denotes the field operator of the Fermi gas. Both species have two internal spin states, which are labelled as $\uparrow$ and $\downarrow$. The single-particle Hamiltonians $h_B$ and $h_F$ are given by 
\begin{align}
h_{B} & =\frac{\left(k-k_{r}\sigma^z_{B}\right)^{2}}{2m_{B}}+\frac{\Omega_{B}}{2}\sigma^x_{B}+ \frac{\delta_{B}}{2}\sigma^z_{B},\label{eq:hB}\\
h_{F} & =\frac{\left(k-k_{r}\sigma^z_{F}\right)^{2}}{2m_{F}}+\frac{\delta_{F}}{2}\sigma^z_{F},\label{eq:hF}
\end{align}
with $\Omega_{B}$ the Raman
coupling strength, $k_{r}$ the Raman recoil momentum, and $\sigma^x_{B,F}$
and $\sigma^z_{B,F}$ the Pauli matrices. We set the two-photon detuning
$\delta_{B}=\delta_{F}=0$ in our discussion. For simplicity, we assume a quasi-one dimensional system with strong transverse confinement. The key result that the fermions experience an effective SO coupling is insensitive to the dimensionality.

The last three terms in Eq. (\ref{eq:H}) describe three types of
two-body interactions, where the Bose-Bose interactions read as 
\begin{equation}
\mathcal{G}_{B}=\int dx\left[g^{B}\left(\rho_{B\uparrow}^{2}+\rho_{B\downarrow}^{2}\right)+2g_{\uparrow\downarrow}^{B}\rho_{B\uparrow}\rho_{B\downarrow}\right],\label{gB}
\end{equation}
with $\rho_{B\sigma}=\left|\psi_{B\sigma}\left(x\right)\right|^{2}$
the spin-$\sigma$ density of the BEC, the Fermi-Fermi interaction
takes the form 
\begin{equation}
\mathcal{G}_{F}=g^{F}\int dx \,\psi_{F\uparrow}^{\dagger}\psi_{F\downarrow}^{\dagger}\psi_{F\downarrow}\psi_{F\uparrow},\label{gF}
\end{equation}
and the Bose-Fermi interactions are given by
\begin{equation}
\mathcal{G}_{BF}=\int dx\left[\gamma\rho_{B}\hat{\rho}_{F}+\beta\text{\ensuremath{\left(\psi_{B\downarrow}^{*}\psi_{B\uparrow}\psi_{F\uparrow}^{\dagger}\psi_{F\downarrow}+h.c.\right)}}\right],\label{eq:gBF}
\end{equation}
where $\rho_{B}=\rho_{B\uparrow}+\rho_{B\downarrow}$ is the density of bosons,
and $\hat{\rho}_{F}=\psi_{F\uparrow}^{\dagger}\psi_{F\uparrow}+\psi_{F\downarrow}^{\dagger}\psi_{F\downarrow}$ is the density operator for fermions.
Here we have assumed that the inter-species density-density interactions are spin-independent, with a single interaction strength $\gamma$, to avoid the proliferation of parameters. The last term in Eq.~(\ref{eq:gBF}) describes the inter-species spin-exchange interaction characterized by the strength $\beta$. We take $L$ to be the length of the system with periodic boundary condition. The number of bosons and fermions are $N_{B,F}$, with the corresponding average densities $n_{B,F}=N_{B,F}/L$, respectively. 

{\em Non-interacting fermions ---} Let us first consider the case with non-interacting fermions, i.e., $g^F=0$. Previous studies of SO coupled BEC have shown that, in the absence of the fermions, the mean-field wave function of the condensate $\Psi_{B}$ can be accurately described by the following 
ansatz:
\begin{equation}
\frac{\Psi_{B}}{\sqrt{n_{B}}}=\left[C_{1}\left(\begin{array}{c}
\cos\theta\\
-\sin\theta
\end{array}\right)e^{ik_{B}x}+C_{2}\left(\begin{array}{c}
\sin\theta\\
-\cos\theta
\end{array}\right)e^{-ik_{B}x}\right],\label{eq:Psi_B}
\end{equation}
where $k_{B}$, $\theta$, $C_{1}$ and $C_{2}$ are variational parameters.
We can restrict $k_{B}\geq 0$ and $\theta\in\left[0,\pi\right]$ without loss of generality, and restrict $C_{1}$ and $C_{2}$
to be real positive numbers, with normalization condition $C_{1}^{2}+C_{2}^{2}=1$, as the relative phase between them will
not affect the total energy. Based on the values of the parameters, three phases
of the SO coupled BEC can be identified: the stripe phase
(ST) with $B\equiv C_{1}C_{2}\neq 0$ and $k_{B}\neq 0$ where the condensate density
profile shows the stripe pattern; the plane-wave phase (PW) with $B=0$
and $k_{B}\neq 0$ where the BEC condenses into a plane-wave state
with finite spin polarization; and the zero-momentum phase (ZM) with $B=0$
and $k_{B}=0$ where the BEC features a smooth density profile with
zero spin polarization. Given the variational ansatz (\ref{eq:Psi_B}), the BEC energy functional $\mathbf{E}_{B}\left(k_{B},\theta,B\right)$,
corresponding to $\int dx\Psi_{B}^{*}h_{B}\Psi_{B}+\mathcal{G}_{B}$,
is given by 
\begin{align}
\frac{\mathbf{E}_{B}\left(k_{B},\theta,B\right)}{N_{B}}= & \frac{k_{B}^{2}+k_{r}^{2}-2k_{r}k_{B}\cos\left(2\theta\right)}{2m_{B}}-\frac{\Omega_{B}}{2}\sin\left(2\theta\right)\nonumber \\
 & -F(B)\cos^{2}\left(2\theta\right)+G_{1}\left(1+2B^{2}\right),\label{eq:E_B}
\end{align}
where we have defined $F(B)=\left(2G_{1}+4G_{2}\right)B^{2}-G_{2}$ and 
$G_{1,2}=n_{B}\left(g^{B}\pm g_{\uparrow\downarrow}^{B}\right)/2$.

\begin{figure}
\includegraphics[scale=0.48]{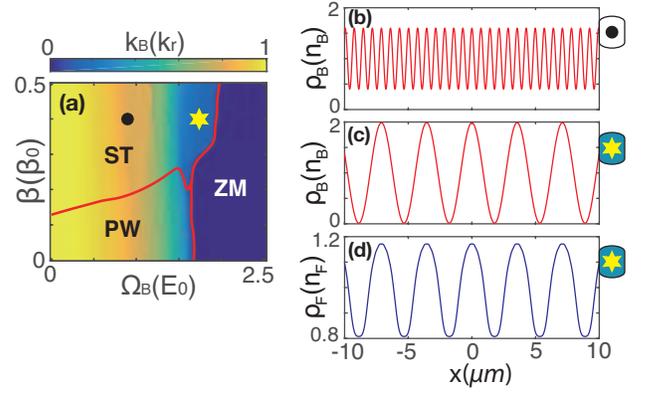}
\caption{(color online) (a) Phase diagram of the BEC characterizing the stripe (ST), plane-wave (PW), and zero-momentum (ZM) phases in the $\Omega_{B}$-$\beta$ plane with $\gamma=0$, where the Fermi-Fermi interaction $g^{F}=0$, the density of bosons $n_{B}=10n_{F}$, and the background color displays the value of $k_{B}/k_{r}$. (b) The boson density profiles for the black dot in (a). (c) and (d) are, respectively, the boson and fermion density profiles for the yellow star in (a). The fermion number is set as $N_{F}=2000$. The mass ratio is taken to be $m_B/m_F=4$. We define the Fermi momentum $k_F = \pi n_F/2$ where $n_F=N_F/L$ is the total fermion density, and $\beta_0=E_0/k_F$ where $E_0=k_F^2/(2m_F)$. The Raman recoil momentum is taken to be $k_{r}=5k_{F}/4$. The condensate interaction strengths are taken to be $g^{B}=6.48*10^{-3}k_{r}/\left(2m_{B}\right)$ and $g_{\uparrow\downarrow}^{B}=2g^{B}$. In (b)-(d), we set $k_{r}=\sqrt{2}\pi/(804.1 {\rm nm})$ \cite{SOCBosonExp5} to convert the length unit to ${\mu}m$.
} \label{fig1}
\end{figure}

The interplay between the condensate and the fermions
is reflected in the $\mathcal{G}_{BF}$ term in Eq. (\ref{eq:gBF}).
We include its effect in an effective fermionic single-particle
Hamiltonian $h_{F}^{\rm eff}$ defined as 
\begin{equation}
\int dx \,\Psi_{F}^{\dagger}h_{F}^{\rm eff}\Psi_{F}=\int dx\Psi_{F}^{\dagger}h_{F}\Psi_{F}+\mathcal{\mathcal{G}}_{BF}\,.
\end{equation}
Since Eq.~(\ref{eq:Psi_B}) is quite general, we assume that the condensate wave function in the presence fermions can still be faithfully represented by Eq.~(\ref{eq:Psi_B}).  
It follows that
\begin{equation}
h_{F}^{\rm eff}\left(k_{B},\theta,B\right)=\frac{\left(k-k_{r}\sigma_{F}^z\right)^{2}}{2m_{F}}+n_{B}\left(\begin{array}{cc}
\gamma V & -\beta M\\
-\beta M^{*} & \gamma V
\end{array}\right),\label{eq:h_f_eff}
\end{equation}
where 
\begin{align}
M & \equiv\frac{\sin\left(2\theta\right)}{2}+B\sin^{2}\theta e^{-2ik_{B}x}+B\cos^{2}\theta e^{2ik_{B}x};\label{eq:M}\\
V & \equiv2B\sin\left(2\theta\right)\cos\left(2k_{B}x\right)+1.\label{eq:V}
\end{align}
The form of $h_{F}^{\rm eff}$ in Eq.~(\ref{eq:h_f_eff}) clearly shows that there is an effective SO coupling in the Fermi gas, which emerges from its interaction with the condensate. Since the two species influence each other, the SO coupling in both components possesses a dynamic nature. Dynamic synthetic gauge field has recently received much attention \cite{dsoc1,dsoc2,dsoc3,dsoc4}. The spinor mixture system thus provides another platform where dynamic SO coupling emerges naturally.

Diagonalizing $h_{F}^{\rm eff}$ gives a set of fermionic single-particle states. Then the total energy of fermions $\mathbf{E}_{F}\left(k_{B},\theta,B\right)$
is obtained by summing up the lowest $N_{F}$ eigenenergies of $h_{F}^{\rm eff}$. The ground state of the mixture is then obtained by minimizing the total energy functional $\mathbf{E}_{B}\left(k_{B},\theta,B\right)+\mathbf{E}_{F}\left(k_{B},\theta,B\right)$ with respect to the variational parameters. In our result, the final values of $\theta$ and $k_{B}$ roughly keep the relation $\cos(2\theta) \approx k_{B}/k_{r}$.

This procedure allows us to present the phase diagram of condensate in the $\Omega_B$-$\beta$ parameter space as shown in Fig.~\ref{fig1}(a), where we take $n_B=10n_F$ and $N_F=2000$. To isolate the effect of the inter-species interactions, we take the density-density interaction strength $\beta=0$. In the absence of fermions, the condensate only possesses two phases, PW and ZM, for $g^B_{\uparrow \downarrow}>g^B$. The transition between them occurs around $\Omega_B= 4k_r^2/(2m_B)$. A notable feature of Fig~\ref{fig1}(a) is that the region with large $\beta$ is dominated by the ST phase. This feature is clearly induced by the fermions. Specifically, the spin-exchange interaction induces an attractive interaction between the two spin components of the condensate, leading to a reduced effective $g^B_{\uparrow \downarrow}$, which favors the ST phase. The background color in Fig.~\ref{fig1}(a) displays the value of $k_B$. In the ST phase, the condensate density profile is given by 
\begin{equation}
\rho_{B}\left(x\right)=n_{B}\left[1+\sin\left(2\theta\right)\cos\left(2k_{B}x\right)\right], 
\end{equation}
with a density modulation whose spatial period is determined by $1/k_B$. One can see that, for a given $\beta$, $k_B$ decreases as $\Omega_B$ increases. Figure~\ref{fig1}(b) and (c) show two condensate density profiles corresponding to the black dot and yellow solid star in (a), respectively. For realistic parameters, the ST phase can possess a spatial period of several microns and a large modulation depth. Such a state can be readily observed using {\em in situ} imaging with today's technology. The density profile for the Fermi gas, corresponding to the yellow solid star, is shown in (d). The two density profiles in (c) and (d) exhibit in-phase modulations.

%
%

Let us now turn to a more in-depth discussion of the properties of fermions in the mixture. When the condensate is in the PW or the ZM phase, we have $B=0$, and the effective single-particle Hamiltonian for fermions $h_F^{\rm eff}$ in Eq.~(\ref{eq:h_f_eff}) is reduced to (after neglecting a term proportional to the constant $n_B$)
\begin{equation}
h_{{\rm F},{\rm PW}}^{{\rm eff}}=\frac{\left(k-k_{r}\sigma_{F}^z\right)^{2}}{2m_{F}}+\frac{\Omega_{F}^{\rm eff}}{2}\sigma_{F}^{x} ,\label{eq:h_f_eff_PW}
\end{equation}
which has the same form as the Hamiltonian of an SO coupled Fermi gas, only that here the SO coupling is not due directly to the Raman lasers, but to the inter-species spin-exchange interaction with an effective Raman coupling strength given by
\begin{equation}
\Omega_{F}^{\rm eff}=-\beta n_{B}\sin\left(2\theta\right).\label{eq:Omega_F_eff}
\end{equation}

\begin{figure}
\includegraphics[scale=0.352]{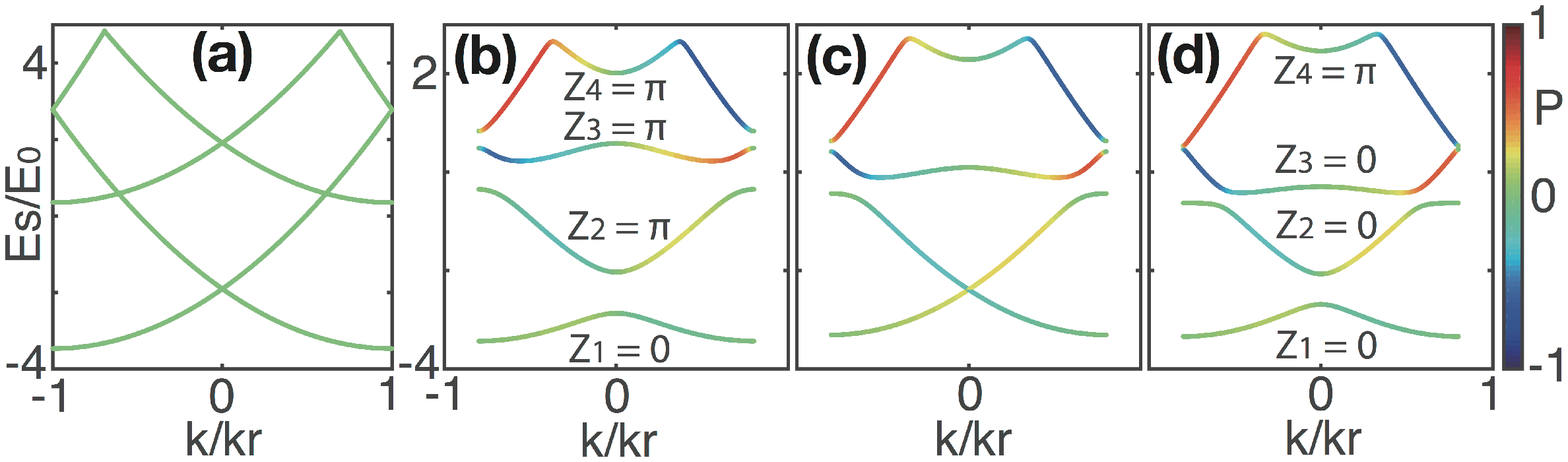}
\caption{(color online) The lowest four energy bands of the non-interacting fermions with $g^{F}=0$, when the condensate is in the ST phase. Here $\beta=0.6\beta_{0}$ in all plots. In (a), $\Omega_B=0$ and $\gamma=0$. For the rest of the plots, $\Omega_{B}=E_{0}$, and $\gamma=0$ (b), 
$\gamma=0.23\gamma_{0}$ (c), and $\gamma=0.4\gamma_{0}$ (d).
The Zak phase for each band is indicated in (b) and (d). The color of the curve denotes the  spin polarization $P=\langle \sigma^z_F \rangle$. 
The other parameters are the same as those in Fig. 1.}
\label{band}
\end{figure}

When the condensate is in the ST phase, we have $B=1/2$, both $V$ and $M$ in Eq.~(\ref{eq:h_f_eff}) exhibit spatial modulations, originated from the density modulation of the condensate. The $V$-term, arising from the inter-species density-density interaction, serves as a lattice potential for the fermions, while the $M$-term, from the spin-exchange interaction, can be regarded as a periodic Raman coupling for the two spin components of the fermions. This situation is analogous to the optical Raman lattice proposed by Liu {\em et al.} \cite{SOCFermionMajorana1, Ramanlattice}, and realized in recent experiments \cite{RLexp1, SOCFermionTopoBand1}. In the Raman lattice setup, the atom experiences an optical lattice potential and a periodic Raman coupling, both originated from the same laser beams. It is shown that the system parameters can be adjusted and induce topological phase transitions. Drawing from this analogy, we also expect topological phases in our system. Figure~\ref{band} displays the lowest four energy bands $E_{s}$ of the effective fermionic single-particle Hamiltonian $h_{F}^{\rm eff}$ in Eq.~(\ref{eq:h_f_eff}), when the condensate is in the ST phase. In all the plots in Fig.~\ref{band}, we fix the value of $\beta$. Figure~\ref{band}(a) is a reference plot where $\Omega_B=0$, hence there is no SO coupling in the system. Here different bands cross each other. The remaining three plots correspond to the same finite value of $\Omega_B$, with varying $\gamma$. In these cases, gaps open up at band crossing points in (a). The color of each band represents the spin polarization $P=\langle \sigma_F^z \rangle$, which can be seen to be momentum-dependent --- a manifestation of the SO coupling. The values of $Z_j$, indicated in (b) and (d), are the Zak phase for each band, defined as \cite{Zak}
\begin{equation}
e^{iZ_{j}}=\stackrel[a=-d]{d}{\prod}w_{j}^{*}\left(k_{a}\right)\cdot w_{j}\left(k_{a+1}\right),\label{eq:Zj}
\end{equation}
where $w_{j}\left(k_{a}\right)$ is the eigenstate of band $j$ and discretized momentum $k_{a}$, restricted in the first Brillouin zone in the range $k_a \in [-k_B, k_B)$, with the additional constraint $w_{j}\left(k_{d+1}\right)=w_{j}\left(k_{-d}\right)$
to form a loop in the calculation of the Zak phase. At a critical value of $\gamma$ shown in Fig.~\ref{band}(c), the lowest two bands crosses each other. When the band reopens at a larger value of $\gamma$, the Zak phase of some of the bands changes its value. Thus the closing and the reopening of the band gap signals a topological transition. Note that Zak phase in topological Bloch bands has been measured in recent cold atom experiments \cite{zakexp}.

{\em Interacting fermions ---} Now let us turn to the situation where the fermions are self-interacting with an attractive $s$-wave interaction strength $g^{F}=-6k_{F}/\left(\pi m_{F}\right)$ in Eq.~(\ref{gF}), which can lead to superfluid pairing. Including Fermi-Fermi interaction greatly complicates the physics in the ST phase \cite{STint}. To keep things relatively simple, we take a large boson density $n_{B}=500n_{F}$, such that in the parameter space we will explore, the bosons are nearly unaffected by the fermions and remain in either the PW or the ZM phase. Under this situation, the effective fermionic single-particle Hamiltonian is given by $h_{\rm F, PW}^{\rm eff}$ in Eq.~(\ref{eq:h_f_eff_PW}). We thus have a system of attractive Fermi gas subjected to SO coupling with an effective Raman coupling strength $\Omega_F^{\rm eff}$ defined in Eq.~(\ref{eq:Omega_F_eff}). The corresponding fermionic system has been studied before \cite{SOCFermionTopoBand6} and is known to support topological superfluid phase. This can be intuitively understood as follows. The SO coupling mixes spin singlet and triplet pairings. The Raman term, which can be regarded as an effective Zeeman field, tends to weaken the singlet pairing. For sufficiently large $\Omega_F^{\rm eff}$, the singlet pairing is suppressed, and the Fermi gas becomes a topological superfluid with effective $p$-wave pairing. This picture is indeed confirmed by our calculation.

\begin{figure}
\includegraphics[scale=0.188]{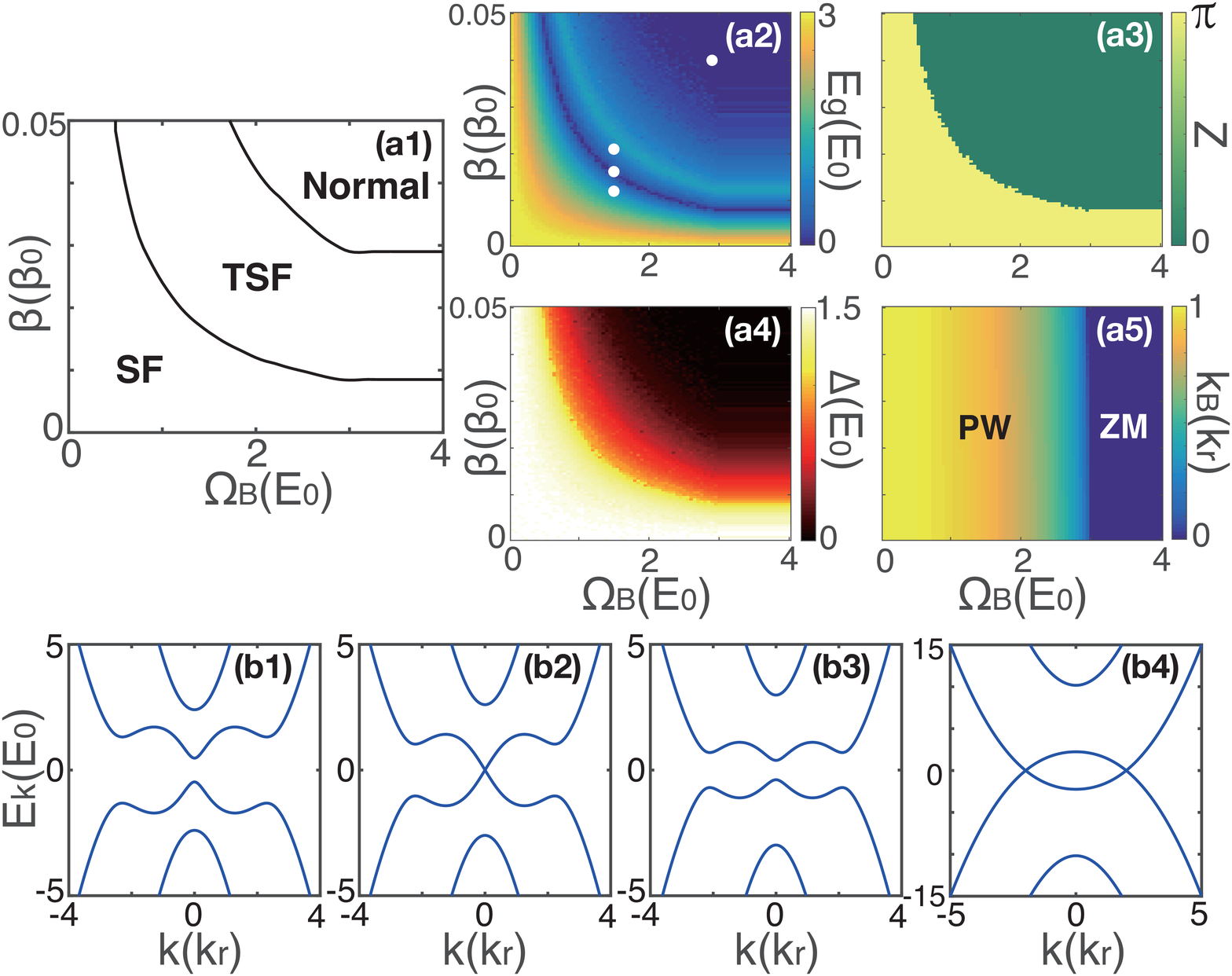}
\caption{(color online) (a1) The phase diagram of the fermions identifying the superfluid (SF), topological
superfluid (TSF), and normal phases in the $\Omega_{B}$-$\beta$
parameter space with attractive Fermi-Fermi interaction strength $g^{F}=-6k_{F}/\left(\pi m_{F}\right)$
and $n_{B}=500n_{F}$. In the same parameter space, we also plot: (a2) the quasi-particle excitation gap $E_{g}$; (a3) the winding number
$Z$; (a4) the Fermi superfluid order parameter $\Delta$; and (a5) the
variational momentum $k_{B}$ of the condensate. The four excitation spectra $E_{k}$ corresponding to the 4 red dots in (a2), from bottom to top, are plotted in (b1) - (b4), respectively. We take
$N_{F}=800$. The other parameters are the same as those
in Fig. 1.}
\label{FF}
\end{figure}

Our calculation proceeds as follows. The bosonic contribution to the total energy functional $\mathbf{E}_{B}\left(k_{B},\theta,B\right)$ still takes the form of Eq.~(\ref{eq:E_B}), only that the parameter $B$ vanishes for the PW and the ZM phase. For the fermionic part, the
thermodynamic grand potential can be written as 
\begin{equation}
\mathbf{P}_{F}\left(\mu,\theta\right)=\int dx\Psi_{F}^{\dagger}\left(h_{{\rm F,PW}}^{{\rm eff}}-\mu\right)\Psi_{F}+\mathcal{\mathcal{G}}_{F},
\end{equation}
where $\mu$ is the chemical potential, and $h_{{\rm F,PW}}^{{\rm eff}}$
and $\mathcal{\mathcal{G}}_{F}$ are defined by Eqs. (\ref{eq:h_f_eff_PW}) and (\ref{gF}), respectively. Note that $h_{{\rm F,PW}}^{{\rm eff}}$
is a function of the variational parameter $\theta$. In our treatment of the fermionic part, we follow the standard mean-field approach introduced for the
single-species interacting fermions as reported in, e.g.,  Refs~\cite{SOCFermionTopoBand6,SOCFermionMajorana4}.
In the mean-field approximation, the Fermi-Fermi interaction term
becomes 
\begin{equation}
\mathcal{\mathcal{G}}_{F}=-\Delta\left[\psi_{F\uparrow}^{\dagger}\left(x\right)\psi_{F\downarrow}^{\dagger}\left(x\right)+{\rm h.c.}\right]-\Delta^{2}/g^{F},
\end{equation}
where the superfluid order parameter is defined as \[\Delta=-g^{F}\left\langle \psi_{F\downarrow}\left(x\right)\psi_{F\uparrow}\left(x\right)\right\rangle \,.\]
In momentum space, the grand potential can be re-written
as 
\begin{equation}
\mathbf{P}_{F}\left(\mu,\theta,\Delta\right)=\frac{1}{2}\sum_{k}C_{k}^{\dagger}M_{k}C_{k}+\sum_{k}\xi_{k}-\frac{L\Delta^{2}}{g^{F}},
\end{equation}
where $C_{k}^{\dagger}=\left[\begin{array}{cccc}
c_{k\uparrow}^{\dagger} & c_{k\downarrow}^{\dagger} & c_{-k\uparrow} & c_{-k\downarrow}\end{array}\right]$ 
and 
\begin{equation}
M_{k}=\left[\begin{array}{cccc}
\xi_{k}+\lambda k & \Omega_{{\rm F}}^{{\rm eff}}/2 & 0 & -\Delta\\
\Omega_{{\rm F}}^{{\rm eff}}/2 & \xi_{k}-\lambda k & \Delta & 0\\
0 & \Delta & -\xi_{k}+\lambda k & -\Omega_{{\rm F}}^{{\rm eff}}/2\\
-\Delta & 0 & -\Omega_{{\rm F}}^{{\rm eff}}/2 & -\xi_{k}-\lambda k
\end{array}\right],
\end{equation}
with $\xi_{k}=k^{2}/\left(2m_{F}\right)-\mu$ and $\lambda=-k_{r}/m_{F}$. Diagonalizing
the matrix $M_{k}$, we can further transform the grand potential to the following form:
\begin{align}
& \mathbf{P}_{F}\left(\mu,\theta,\Delta\right)=\frac{1}{2}\sum_{k}(E_{k1}\alpha_{k1}^{\dagger}\alpha_{k1}+E_{k2}\alpha_{k2}^{\dagger}\alpha_{k2}\\
& +E_{k3}\alpha_{k3}\alpha_{k3}^{\dagger}+E_{k4}\alpha_{k4}\alpha_{k4}^{\dagger})+\sum_{k}\xi_{k}-\frac{L\Delta^{2}}{g^{F}},\nonumber 
\end{align}
where $\alpha_{k1}$, $\alpha_{k2}$, $\alpha_{k3}$, and $\alpha_{k4}$
are quasi-particle elementary excitation operators with the symmetry
$E_{k4}=-E_{-k1}$ and $E_{k3}=-E_{-k2}$. The two
positive excitation branches are given by 
\begin{equation}
E_{k1,2}=\left(\xi_{k}^{2}+\eta_{k}+\Delta^{2}\pm\sqrt{4\eta_{k}\xi_{k}^{2}+\left(\Omega_{{\rm F}}^{{\rm eff}}\right)^{2}\Delta^{2}}\right)^{\frac{1}{2}},\label{eq:Ek12}
\end{equation}
with $\eta_{k}=\left(k_{r}k/m_{F}\right)^{2}+\left(\Omega_{{\rm F}}^{{\rm eff}}\right)^{2}/4$. The ground state of the fermions is considered
to be the quasi-particle vacuum, with the corresponding ground-state grand
potential given by 
\begin{equation}
\mathbf{P}_{F}\left(\mu,\theta,\Delta\right)=-\frac{1}{2}\sum_{k}\left(E_{k1}+E_{k2}\right)+\sum_{k}\xi_{k}-\frac{L\Delta^{2}}{g^{F}},
\end{equation}
where the anti-commutation relations of $\alpha_{k1}$, $\alpha_{k2}$,
$\alpha_{k3}$, and $\alpha_{k4}$ have been considered. Note that
$\mathbf{P}_{F}\left(\mu,\theta,\Delta\right)$ is a functional of
three undetermined variational parameters $\mu$, $\theta$, and $\Delta$.

The ground state of the whole mixture is obtained through the minimization
of $\mathbf{E}_{B}\left(k_{B},\theta,0\right)+\mathbf{P}_{F}\left(\mu,\theta,\Delta\right)$
with respect to the variational parameters $k_{B}$, $\theta$, and
$\Delta$, where the constraint  $N_{F}=-\partial\mathbf{P}_{F}\left(\mu,\theta,\Delta\right)/\partial\mu$
is imposed to fix the number of fermions. Actually, the minimization
of $\mathbf{E}_{B}\left(k_{B},\theta,0\right)$ with respect to $k_{B}$
leads to the rigorous relation $\cos\left(2\theta\right)=k_{B}/k_{r}$,
and hence we only need to deal with $\theta$ and $\Delta$ in the
numerical minimization. The converged results of $k_{B}$, $\theta$,
$\Delta$, and $\mu$ are obtained in the thermodynamic limit with
$N_{B},N_{F},L\rightarrow\infty$ while keeping $n_{B}=N_{B}/L$ and $n_{F}=N_{F}/L$ finite.

Our results are summarized as follows. Figure~\ref{FF}(a1) represents the zero temperature phase diagram of the Fermi gas in the $\Omega_B$-$\beta$ space. It shows three phases: the non-topological superfluid (SF), the topological superfluid (TSF), and the normal phase. The first two phases feature finite superfluid order parameter $\Delta$, whereas $\Delta$ vanishes in the normal phase, as shown in Fig.~\ref{FF}(a4). The quasi-particle excitation gap $E_{g}$ is finite in the SF and the TSP phases, except at the boundary of these two phases, as shown in Fig.~\ref{FF}(a2), where $E_g$ vanishes as expected for topological phase transition. Several examples of the quasi-particle excitation spectra $E_{k}$ at various phases are displayed in Fig.~\ref{FF}(b1)$\sim$(b4). 

The topological phase transition can be further confirmed by the winding number $Z$, which is defined through a loop connecting the two positive excitation branches at infinitely large $k$, and can be calculated as \cite{SOCFermionMajorana4}
\begin{equation}
e^{iZ}=\stackrel[a=-b]{b}{\prod}\left[w_{1}^{*}\left(k_{a}\right)\cdot w_{1}\left(k_{a+1}\right)\right]\left[w_{2}^{*}\left(k_{a}\right)\cdot w_{2}\left(k_{a+1}\right)\right],\label{eq:Z_sf-1}
\end{equation}
where $w_{1}\left(k_{a}\right)$ and $w_{2}\left(k_{a}\right)$ are
the quasiparticle eigenstates corresponding to the two positive excitation
branches $E_{k1}$ and $E_{k2}$ in Eq.~(\ref{eq:Ek12}), with the
quasi-momentum defined as $k_{a}=2\pi a/L$ with the integer $a\in\left[-b,b\right]$
where $b$ is a sufficiently large numerical cut-off.
We define $w_{1}\left(k_{b+1}\right)=w_{2}\left(k_{-b}\right)$
and $w_{2}\left(k_{b+1}\right)=w_{1}\left(k_{-b}\right)$ to connect
the two positive excitation branches and form the loop in calculating
the winding number. Note that here $k_{a}$ in (\ref{eq:Z_sf-1})
is no longer restricted in the first Brillouin zone. As shown in Fig.~\ref{FF}(a3), $Z$ jumps from $\pi$ to 0 when entering from SF to TSF. Finally, through combining the result of $k_B$ in Fig.~\ref{FF}(a5), the definition of $\Omega_F^{\rm eff}$ in Eq.~(\ref{eq:Omega_F_eff}), and the relation $\cos (2\theta)=k_B/k_r$, we can see that the TSF phase does correspond to the larger $\Omega_F^{\rm eff}$, and hence our result agrees with our previous picture.


{\em Summary ---} In summary, we have investigated a system of Bose-Fermi spinor mixtures. The bosons form a condensate that subjected to the Raman-induced SO coupling, while the fermions are not coupled to the Raman lasers, but interact with the bosons via the density-density and/or spin-exchange interaction. We show that the spin-exchange interaction makes the fermions experiencing an effective SO coupling, without suffering Raman-induced heating. This could pave a new way towards the first realization of SO coupled fermionic superfluids, which can be made to be topological with a proper choice of parameters. The interplay between the bosons and fermions also has an interesting effect on the former: the Bose-Fermi interaction favors the condensate to be in the ST phase, with an interaction-dependent spatial modulation period. With realistic parameters, the spatial modulation period can be as large as several microns, making the ST phase readily observable with the {\em in situ} imaging technique. This provides a significant advantage in both the realization and the observation of the ST phase. The phenomena described above arise due to the emergent and dynamic nature of the SO coupling in spinor mixtures.

Our proposal does not require any new experimental techniques beyond those that have already been demonstrated in the lab. In particular, both the Bose-Bose spinor mixtures \cite{DJWang2} and the Bose-Fermi superfluid mixtures \cite{BFmixSuperfluid1,BFmixSuperfluid2,BFmixSuperfluid3,BFmixSuperfluid4} have been realized. The density-density interaction strengths have been routinely tuned via Feshbach resonance \cite{Feshbach2,Feshbach3}. Recent works have also shown that the spin-exchange interaction can be tuned to some extent \cite{SpinExchInt1,SpinExchInt2,SpinExchInt3, SpinExchInt4,SpinExchInt5}. We therefore expect that our proposal will inspire more works, both theoretical and experimental, on SO coupled quantum gas spinor mixture, providing a new and unique platform to explore the physics of SO coupling.  

We acknowledge discussions with Xiong-Jun Liu, Randall G. Hulet, and Peng Zhang. HP is supported by the US NSF and the Welch Foundation (Grant No. C-1669), LC by the NSF of China (Grant No. 11804205), HH and XJL by the Australian Research Council (Grant No. DP170104008).


\begin{thebibliography}{10}
\bibitem{STint} When the condensate is in the ST phase, fermions experiences a band structure as shown in Fig.~\ref{band}. In the presence Fermi-Fermi interaction, it becomes a quite difficult problem to treat the bosons and fermions self-consistently for such an inhomogeneous system. We plan to address this problem in the future.
\end{thebibliography}
\end{document}